\documentclass{elsart3}

\usepackage{natbib}

\usepackage{graphics}
\usepackage{graphicx}
\usepackage{epsfig}

\usepackage{amssymb}

\newcommand{\msun}{\,\hbox{M$_{\odot}$}}
\newcommand{\lsun}{\,\hbox{L$_{\odot}$}}
\newcommand{\kms}{\,\hbox{\hbox{km}\,\hbox{s}$^{-1}$}}

\newcommand{\mbh}{\,\hbox{M$_{\rm BH}$}}
\newcommand{\msigma}{\,\hbox{M$_{\rm BH}-\sigma$}}

\begin{document}

\begin{frontmatter}


\title{Probing for evolutionary links between local ULIRGs and QSOs using NIR 
spectroscopy}

\author{K. M. Dasyra, L. J. Tacconi, R. I. Davies, R. Genzel, D. Lutz}
\address{Max-Planck-Institut f\"ur extraterrestrische Physik,
85741 Garching, Germany}
\ead{dasyra@mpe.mpg.de, linda@mpe.mpg.de, davies@mpe.mpg.de, genzel@mpe.mpg.de, lutz@mpe.mpg.de}

\author{T. Naab}
\address{Univesit\"atssternwarte, Scheinerstr. 1, 81679, M\"unchen, Germany}
\ead{naab@usm.uni-muenchen.de}

\author{D. B. Sanders}
\address{Institute for Astronomy, University of Hawaii, 2680 Woodlawn
  Drive, Honolulu, HI 96822, USA}
\ead{sanders@ifa.hawaii.edu}

\author{S. Veilleux}
\address{Department of Astronomy, University of Maryland, College
  Park, MD 20742, USA}
\ead{veilleux@astro.umd.edu}
\author{A. J. Baker}
\address{Jansky Fellow, National Radio Astronomy Observatory}
\address{Department of Astronomy, University of Maryland, College
  Park, MD 20742, USA}
\ead{ajb@astro.umd.edu}

\begin{abstract}
We present a study of the dynamical evolution of Ultraluminous Infrared 
Galaxies (ULIRGs), merging galaxies of infrared luminosity $>10^{12}$ \lsun. 
During our Very Large Telescope large program, 
we have obtained ISAAC near-infrared, high-resolution spectra of 54 ULIRGs (at 
several merger phases) and 12 local Palomar-Green QSOs, to investigate 
whether ULIRGs go through a QSO phase during their evolution. One possible 
evolutionary scenario is that after nuclear coalescence, the black hole 
radiates close to Eddington to produce QSO luminosities. The mean stellar 
velocity dispersion that we measure from our spectra is similar 
($\sim$160 km/s) for 30 post-coalescence ULIRGs and 7 IR-bright QSOs. 
The black holes in both populations have masses of order 10$^7$-10$^8$ 
\msun\ (calculated from the relation to the host dispersion) 
and accrete at rates $>0.5$ Eddington. Placing ULIRGs 
and IR-bright QSOs on the fundamental plane of early-type galaxies shows 
that they are located on a similar region (that of moderate-mass 
ellipticals), in contrast to giant ellipticals and radio-loud QSOs. 
While this preliminary comparison of the ULIRG and QSO host kinematical
properties indicates that (some) ULIRGs may undergo a QSO phase in their
evolutionary history before they settle down as ellipticals, further data 
on non-IR excess QSOs are necessary to test this scenario.
\end{abstract}

\end{frontmatter}

\section{Introduction}
\label{intro}
In the local Universe, the best laboratories for studying violent merging 
events (considered key-mechanisms in driving galaxy evolution) are 
the ultraluminous infrared galaxies (ULIRGs). Several studies indicate that 
ULIRGs transform gas-rich disks into moderate mass ellipticals through 
merger-induced dissipative collapse \citep{kor92,mihos96}. Photometric
analysis of ultraluminous merger remnants by \citet{veilleux02} and 
\citet{sco00} indicates that most of the sources are are well-fit by an 
elliptical-like r$^{1/4}$ light profile. Specrtoscopic analysis by 
\citet{genzel01} and \citet{tacconi02} indicates that their stellar kinematics
resemble those of dispersion-supported systems.

While the end products of galactic mergers can be considered understood, 
several details of the merging process (often related to the physics 
of the gas) are still very uncertain, even in the local Universe. 
A plethora of numerical models \citep{mihos96,springel05} and 
observations \citep{sami96,kim02} indicates that major 
starburst episodes occur after the first galactic encounter and can be 
present after nuclear coalescence, before complete relaxation sets in.
According to \citet{dimatteo}, the main accretion event happens after
the coalescence. However, the relative contributions of starburst and AGN 
to the infrared (IR) emission during each phase of a gas-rich merger are still 
unclear. Whether ULIRGs may go through a QSO phase after the nuclear 
coalescence also needs to be confirmed. 

One way to investigate the physical details of mergers is to determine the 
evolution of the ULIRG kinematic properties as a function of time. Hence, 
we obtained high-resolution, H- and K- band ISAAC spectroscopic data for 54 
ULIRGs (spanning a wide range of merger phase) during our European Southern 
Observatory Very Large Telescope (VLT) large program (ID 171.B-0442). Of 
these sources (at $0.018<z<0.268$), 23 are binary (not fully merged) ULIRGs 
and 30 are (relaxed) remnants. We extract  
the stellar dispersion $\sigma$ and rotational $V_{\rm rot}$ velocity from 
the CO rovibrational bandheads that appear in our spectra using the Fourier 
correlation quotient technique \citep{bender90}; this method provides the 
line-of-sight (LOS) broadening function with which a stellar template has to 
be convolved to produce the observed spectrum.

\section{Dynamics of ultraluminous infrared mergers}
\subsection{Pre-coalescence phase: the conditions that trigger  
ultraluminous emission}
\label{pre}
We derive the kinematics of each individual component of the binary systems,
to investigate the conditions under which ultraluminous infrared activity
is triggered. From the stellar dispersion and rotational velocity, we 
compute the masses of the merging galaxies. We find that that ultraluminous 
luminosities are
are mainly generated by almost equal-mass mergers; the average mass ratio 
of the binary ULIRGs is 1.5:1 and 68\% of these sources are 1:1 encounters 
\citep{dasyra05}. Statistically, our result is in agreement with luminosity 
ratios inferred from imaging studies \citep[e.g.][]{kim02}. In individual 
cases though, luminosity does not always accurately trace mass due to stellar 
population and extinction effects.

To investigate whether dynamical heating (mainly of the smaller companion)
can skew our results, we re-ran the simulations of \citet{naab03} with an
$N$-body-plus-SPH code after replacing 10\% of the stellar mass with 
isothermal (10$^4$ K) gas. Following the stellar dispersion as a function 
of time, we found that 3:1 mergers may appear to be 2:1 at nuclear 
separations as small as those we observe (7.2 kpc on average). Since
only a minority of our sources belong to these merger categories, our 
conclusions are stable against dynamical heating. Mergers of mass 
ratio $>$4:1 appear not to drive enough gas to the center of the merger 
to generate ultraluminous IR emission.

\subsection{Post-coalescence phase: the evolution of the host properties and 
the end-products of ultraluminous mergers}
\label{post}
By studying the kinematics of ULIRGs at merger phases that follow  
nuclear coalescence (remnants) and comparing them to the kinematics
of the binary ULIRGs, we find that the mean stellar dispersion increases 
during the coalescence phase by 15 \kms\ (namely from 142 to 157 \kms). 
Although this increase is significant (within the 0.05 significance level 
of our Monte Carlo simulations), it constitutes only a lower limit on
the dynamical heating that the merging galaxies undergo. This is primarily 
due to the fact that the ultraluminous phases are short compared to the total
merger timescales \citep[e.g.][]{mihos99}.

\begin{figure*} 
\begin{center}
\includegraphics[width=9cm]{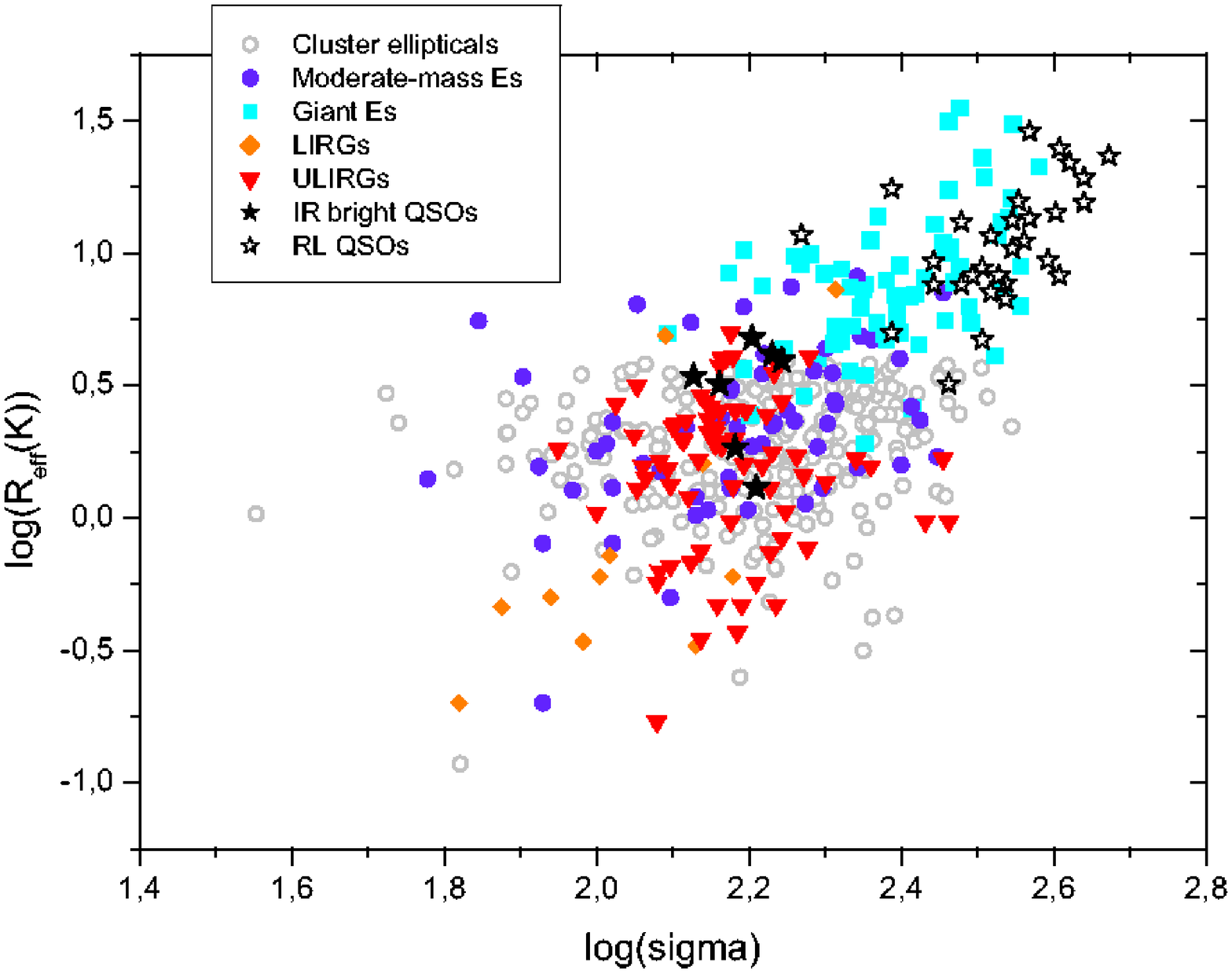}
\end{center} 
\caption{The fundamental plane of early-type galaxies ($_{\rm eff}-\sigma$
projection). The giant boxy and intermediate-mass disky Es 
(boxes and filled circles respectively) are from \citet{bender92} and 
\citet{faber97}. Cluster Es (open circles) are taken from \citet{pahre}
and LIRGs (diamonds) from \citet{shier98} and \citet{james99}. The ULIRGs 
and the IR-bright QSOs of our program appear as triangles and filled stars 
respectively. The RL/RQC QSOs of \citet{dunlop03} are open stars. Their
black hole masses are converted into dispersions using the \msigma\ relation
\citep{tremaine02}. 
\label{fig:fpp} } 
\end{figure*}

The ratio between rotational and dispersion velocity of the ULIRG remnants, 
which equals 0.31 and increases to 0.62 when inclination effects are taken 
into account, indicates that the remmants are dispersion-supported systems, 
resembling ellipticals (Es). 
To investigate what type of ellipticals ultraluminous mergers will form, 
\citet{genzel01} suggested that ULIRGs need to be placed on the fundamental 
plane (FP) of early-type galaxies \citep{djoda}. 
In Fig.~\ref{fig:fpp} we present the effective radius- host dispersion
($R_{\rm eff}-\sigma$) projection of the plane, constructed from our data 
for 54 ULIRGs (triangles). Giant (boxy-isophotal profile) Es are shown in 
boxes, moderate-mass (disky-isophotal profile) Es in filled circles, and 
further cluster Es in open circles (see caption for original references). 
Some luminous infrared galaxies (LIRGs; 
$10^{11} L_{\odot} < L < 10^{12} L_{\odot}$) are presented in diamonds.
The fact that ULIRGs clearly populate the moderate-mass ($\sim$10$^{11}$ 
M$_{\odot}$) Es part of the FP suggests that these two populations are linked, 
while giant Es probably have a different formation history.

From the dispersion velocity of the stars in the bulge of our ULIRG remnants, 
we compute their BH mass $M_{BH}$ using the $M_{BH}-\sigma$ relation 
\citep[e.g.][]{tremaine02}. To test whether the \msigma\ relation can be 
applied at the late ultraluminous phases of a merger, we ran our simulations 
(presented in Sect.~\ref{pre}) and followed the host dispersion and the
gas inflowing to the center of the simulation as a function of time.
We found that by the time the nuclei coalesce, these quantities already 
scale linearly as long as the amount of gas that accretes onto the black 
hole from its surroundings (center of the simulation) is roughly constant 
with time \citep[e.g.][]{dimatteo}. The application of the \msigma\ relation 
to the remnants yields black hole masses of the order $10^7-10^8$ \msun. 

The Eddington efficiency $\eta_{\rm Edd}$ (the ratio between the Eddington 
and the dynamical \mbh) of the ULIRG remnants is calculated by assigning 50\% 
of the IR luminosity to the AGN. This statistical assumption follows 
\citet{genzel98}, who found that some ULIRGs are largely starburst- while 
others are AGN- powered, and may cause a few individual sources to accrete 
at super-Eddington rates. If we assign $\eta_{\rm Edd}$=1 to these sources,
the mean Eddington efficiency is 0.5. Such high $\eta_{\rm Edd}$ values may be 
an observational confirmation of the theoretical predictions of 
\citet{springel05} and \citet{dimatteo}. These authors suggest that after 
nuclear coalescence, the gas infall to the center of the system is so 
high that the AGN may accrete at near-Eddington rates.


\section{Dynamics of Palomar-Green QSO hosts and their relation to mergers}
\label{qso}
An evolutionary scenario for the late phases of gas-rich mergers, originally 
based on that of \citet{sanders88}, is as follows: after coalescence, 
the IR emission arising from the nuclear starburst and AGN-surrounding dust 
is strong enough to reach QSO-like luminosities. However, as the dust and 
gas start clearing out from the nuclear region due to AGN winds and supernova 
feedback, the system  goes through a short (up to ~10$^8$ yrs) optically 
bright phase, before further accretion and star formation are finally 
prevented \citep{springel05}. 

To test this scenario we obtained spectroscopic (VLT ISAAC-PI Tacconi, 
SPITZER IRS-PI Veilleux) and imaging (HST NICMOS-PI Veilleux) data for a 
(small) sample of Palomar-Green QSOs \citep{schmidt} as part of our QUEST
(QSO/ULIRG evolutionary study) project. The VLT sample consists of 12 
sources, most of which are IR-bright (ratio of integrated IR to big blue 
bump luminosity $> 0.46$; Surace et al., 2001). If the above evolutionary
scenario is realistic, the IR-bright sources are transitional objects
between ULIRGs and (optically-selected) QSOs. 

The HST NICMOS imaging analysis of 7 IR-bright QSOs by \citet{veilleux06} 
indicates that their hosts have similar H-band host magnitudes to those of 
ULIRGs. In a similar vein, \citet{canalizo} found that the (optical) spectra 
of most of their IR-bright QSOs (selected from the IRAS color-color diagrams) 
are well fitted by an old population and a recent ($\lesssim 3 \times 10^8$ 
yrs) strong starburst component.

From our high-resolution, large-collecting area data (obtained under
excellent seeing conditions $\lesssim 0.5$), we succeeded in extracting host 
dispersions of local QSOs from NIR CO bandheads. The average dispersion of 
the IR-bright QSOs is (so far) similar to that of ULIRG remnants ($\sim$ 160 
\kms). Thus, their 
black hole and host masses are also of comparable size (10$^7$-10$^8$ and 
10$^{10}$-10$^{11}$ \msun\ respectively). Since IR-bright QSOs have similar 
luminosities ($\gtrsim 10^{12}$ \lsun) to ULIRGs, their accretion rates are 
also high ($\sim0.5$ Eddington). For the few optically selected sources 
observed so far, the scatter in the dynamical properties is higher. 

\begin{figure*}
\begin{center}
\includegraphics[height=6.5cm]{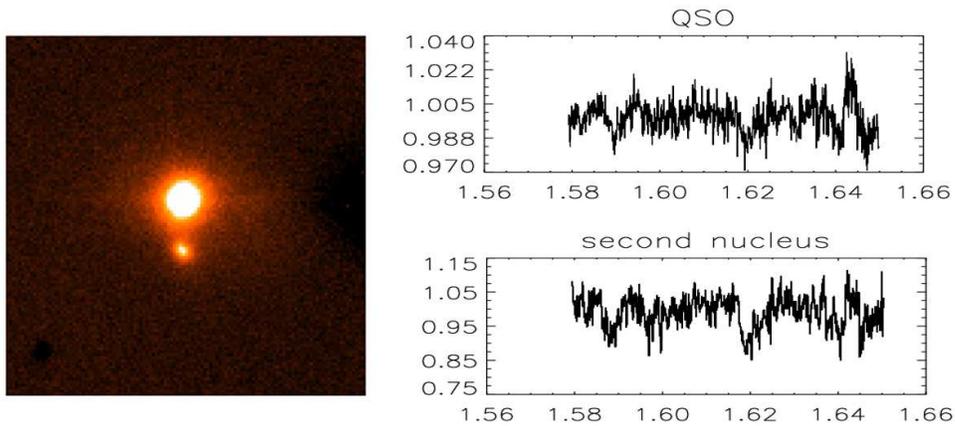}
\end{center}
\caption{PG 1426+015- On the left panel the acquisition image with the
two nuclei is shown. On the right panel, the NIR spectra are presented;
the on-source integration time on the QSO is $\sim$ 2 times deeper than 
that on the secondary nucleus.
\label{fig:pg1426}
}
\end{figure*}

In one of the 12 QSOs targeted by our large program, PG1426+015, two
nuclei separated by $\sim$4kpc appear in the acquisition image (also seen in 
HST NICMOS data; Schade et al., 2000). Our NIR spectroscopy indicates that the 
nuclei are at the same redshift. The optical spectrum of the bright component
has very broad emission lines \citep{kaspi00}, verifying that the emission
originating from this component can be attributed to an AGN.
Calculation of the IR luminosity of the system (from IRAS fluxes) indicates 
that it is a LIRG. The spectra of the QSO and the secondary nucleus are 
presented in Fig.~\ref{fig:pg1426}. The second nucleus does not show any
indications of strong AGN continuum and resembles those of the binary
ULIRGs; therefore we believe that it is starburst dominated. This system 
constitutes a good example of how mergers can simultaneously trigger
strong accretion and starburst events. These findings contradict the 
aforementioned scenario in that the QSO phase is reached already before  
nuclear coalescence.

The IR-bright QSOs seem to differ from local QSOs that host even more massive 
black holes, such as the $0.1<z<0.25$ radio-loud (RL) sources or their 
radio-quiet optical counterparts (RQC) of \citet{dunlop03}. The RL/RQC QSOs 
have black hole masses of order 10$^8$ and 10$^9$ \msun\ that accrete on 
average at rates $\sim$ 0.05 Eddington, and that are located in $~5$ times
more massive and (K-band) luminous hosts.  The positions of IR-bright and
RL/RQC QSOs on the fundamental plane can be clearly distinguished (see 
Fig.~\ref{fig:fpp}); like giant Es vis-a-vis ULIRGs, the RL/RQC QSOs probably 
have a different formation history from the IR-bright population.

Our preliminary results seem to indicate evolutionary links between 
ULIRGs and IR-bright QSOs; however, the fact that some IR QSOs have
prominent spiral hosts (Surace et al., 2001; Veilleux et al., 2006) implies 
that they may have a 
minor-merger origin (different from that of ULIRGs in terms of the mass ratio
of the merging galaxies). Furthermore, IR-bright QSOs are only a small
and not necessarily representative part of the PG poopulation. We need to 
enrich our sample with optically selected QSOs in order to derive conclusions 
about the optically bright phase of this scenario.

\section{Conclusions} 

From the high-resolution NIR spectrocopic data we have obtained for local 
ULIRGs and QSOs, we have found that
\begin{itemize}
\item
For IR luminosity $>10^{12}$ \lsun\ to be triggered during a gas-rich merger,
encounters of comparable mass galaxies are (typically) required. 
\item
Evolution (increase) of the host dispersion is observed as the merger
advances from pre- to post- coalescence.
\item
The merger remnants resemble moderate-mass ($\sim$10$^{10}$-10$^{11}$ \msun)
ellipticals. The black holes they host are of the order 10$^{7}$-10$^{8}$ 
\msun\ and, on average, accrete at high Eddington rates ($\ge$0.5).
\item
The IR bright QSO dispersions and black hole masses, being of the order 
10$^7$-10$^8$ \msun, resemble those of ULIRG remnants and indicate a possible 
link between the two populations. Our IR-bright sources differ from QSOs that 
host supermassive black holes and accrete at low rates.
\item
Imaging and spectroscopy of PG 1426+015 show that it is a binary system of 
nuclear separation $\sim$4 kpc. Already at this early merger phase, one of 
the components is a powerful QSO (with strong dust continuum).
\end{itemize}



\newpage

\end{document}